\documentclass[11pt]{article}
\usepackage{graphicx} 
\usepackage{geometry}
\usepackage{amsmath, amssymb}
\usepackage{bm}
\usepackage{hyperref}
\usepackage{setspace}
\usepackage{graphicx} 
\usepackage{subcaption}
\usepackage{lineno}
\usepackage{float}
\usepackage{booktabs}
\usepackage{threeparttable}
\usepackage{algorithm}
\usepackage{cleveref}
\usepackage{booktabs,threeparttable}
\usepackage{algpseudocode}
\usepackage{authblk}

\newcommand\BibTeX{{\rmfamily B\kern-.05em \textsc{i\kern-.025em b}\kern-.08em
T\kern-.1667em\lower.7ex\hbox{E}\kern-.125emX}}
\usepackage[super,sort&compress]{natbib}

\title{\huge Uncovering Treatment Effect Heterogeneity in Pragmatic Gerontology Trials}
\author[1,2,3]{Changjun Li}
\author[1, 4]{Heather Allore}
\author[5, 6]{Michael O. Harhay}
\author[1,3,7]{Fan Li}
\author[1,2,3,7]{Guangyu Tong\thanks{Correspondence: Guangyu Tong, PhD, 135 College St, \#234, New Haven, CT 06510, USA. Tel: +1 203-785-5553; Email: \texttt{guangyu.tong@yale.edu}.}}

\affil[1]{Department of Biostatistics, Yale School of Public Health, New Haven, CT, USA}

\affil[2]{Cardiovascular Medicine Analytics Center, Yale School of Medicine, New Haven, CT, USA}

\affil[3]{Center for Methods in Implementation and Prevention Science, Yale
School of Public Health, New Haven, CT, USA}

\affil[4]{Section of Geriatrics, Department of Internal Medicine, Yale School of Medicine, New Haven, CT, USA}

\affil[5]{Palliative and Advanced Illness Research (PAIR) Center, Perelman School of Medicine, University of Pennsylvania, Philadelphia, PA, USA}

\affil[6]{Center for Clinical Trials Innovation, Department of Biostatistics, Epidemiology \& Informatics, Perelman School of Medicine, University of Pennsylvania, Philadelphia, PA, USA}

\affil[7]{Section of Cardiovascular Medicine, Department of Internal Medicine, Yale School of Medicine, New Haven, CT, USA}

\date{}

\begin{document}

\maketitle


\begin{abstract}
Detecting heterogeneity in treatment response enriches the interpretation of gerontologic trials. In aging research, estimating the intervention's effect on clinically meaningful outcomes poses analytical challenges when outcomes are truncated by death. For example, in the Whole Systems Demonstrator trial, a large cluster-randomized study evaluating telecare among older adults, the overall effect of the intervention on quality of life was found to be null. However, this marginal intervention estimate obscures potential heterogeneity of individuals responding to the intervention, particularly among those who survive to the end of follow-up. To explore this heterogeneity, we adopt a causal framework grounded in principal stratification, targeting the Survivor Average Causal Effect (SACE)—the treatment effect among “always-survivors,” or those who would survive regardless of treatment assignment. We extend this framework using Bayesian Additive Regression Trees (BART), a nonparametric machine learning method, to flexibly model both latent principal strata and stratum-specific potential outcomes. This enables the estimation of the Conditional SACE (CSACE), allowing us to uncover variation in treatment effects across subgroups defined by baseline characteristics. Our analysis reveals that despite the null average effect, some subgroups experience distinct quality of life benefits (or lack thereof) from telecare, highlighting opportunities for more personalized intervention strategies. This study demonstrates how embedding machine learning methods, such as BART, within a principled causal inference framework can offer deeper insights into trial data with complex features including truncation by death and clustering—key considerations in analyzing pragmatic gerontology trials.
\end{abstract}

\noindent\textbf{Keywords:} Whole Systems Demonstrator Telecare Questionnaire Study; gerontology trials; cluster randomized trial; pragmatic trials; principal stratification; machine learning; Bayesian inference; survivor average causal effect

\section{Introduction}

Pragmatic trials are a cornerstone of gerontologic research, providing rigorous yet policy-relevant evidence on interventions that aim to support independence, quality of life, and health in late life.\cite{roland1998understanding, resnick2022pragmatic} Their strengths lie in large, heterogeneous populations and real-world delivery, often using cluster-randomized designs. \cite{eldridge2012practical} However, these very characteristics can complicate the inference on the intervention performance: treatment effects averaged across diverse participants may appear modest or null, even when meaningful benefits (or harms) exist for specific subgroups. Conventional subgroup analyses, typically prespecified along a few baseline dimensions, are rarely sufficient to detect more complex forms of heterogeneity. As a result, valuable information may remain hidden, and seemingly “negative” trials risk being dismissed despite containing signals that could guide intervention targeting and future research. Therefore, machine learning and other data-driven approaches are increasingly valuable for estimating individualized treatment effects and detecting complex, multidimensional patterns of heterogeneity that conventional subgroup analyses may overlook \cite{kent2018personalized, halvorsrud2007conceptualization}. As one prominent example, parallel advances in machine learning for causal inference have shown that Bayesian Additive Regression Trees (BART) can flexibly learn complex regression surfaces with built-in regularization \cite{chipman2010bart}, and have recently been applied to treatment effect heterogeneity detection in clinical trials. \cite{chen2024bayesian, hu2021estimating, ghazi2025treatment}.

Yet, in gerontologic trials, these opportunities come with unique complications. Quality of life (QoL) outcomes that are among the most clinically meaningful endpoints in late-life intervention studies are a prime example. \cite{halvorsrud2007conceptualization, netuveli2008quality} Measures such as EQ-5D or SF-12 are widely used to capture well-being, functional status, and independence—dimensions of aging that survival alone cannot reflect. \cite{group1990euroqol, ware199612} However, QoL outcomes face a special challenge: they are undefined for participants who die before follow-up, or who deteriorate so severely that reporting is impossible. This phenomenon, known as truncation by death, is especially common in gerontology trials, where mortality is non-negligible during extended follow-up. \cite{rubin2006causal} Because participants who die may differ systematically from those who survive, and because interventions themselves may influence participant survival, analyses restricted to observed survivors risk introducing survivor bias and yield estimates that are difficult to interpret causally. One principled solution comes from principal stratification, which defines causal effects within strata of participants classified by their potential survival under both treatment and control.\cite{frangakis2002principal,zhang2003estimation} In this framework, the survivor average causal effect (SACE) focuses on the “always-survivors”—those who would survive regardless of assignment—for whom both counterfactual QoL outcomes are well defined. Extending this idea to covariate conditional effects (CSACE) enables the investigation of how survivor responses vary across baseline characteristics, uncovering heterogeneity in a way that standard subgroup analyses cannot.\cite{chen2024bayesian}

The death truncation challenge is compounded by the fact that many pragmatic trials in gerontology adopt cluster randomization, either at the practice, facility, or community level, to reduce contamination, facilitate implementation, or reflect the way interventions are delivered in routine care. \cite{eldridge2012practical,murray1998design} Such designs are especially common when evaluating service models, long-term care interventions, or technologies that naturally operate at a group level. For example, the Whole Systems Demonstrator trial of telecare randomized at the general practice level, \cite{steventon2012effect,hirani2014effect} while cluster-randomized or cluster-stepped wedge designs have also been used in fall prevention, dementia care, and palliative interventions in older populations (e.g., \cite{bhasin2020randomized,lamb2020screening}). Compared with individually randomized trials, CRTs introduce additional sources of correlation and heterogeneity, requiring methods that account for within-cluster dependence. A similar issue can arise in another type of closely related design in pragmatic trials called individually randomized group treatment trials (IRGTs), and a description of IRGT can be found elsewhere. \cite{tong2024designing} Recent work has estimated SACE (and covariate conditional extensions, CSACE) in CRTs using parametric principal strata and outcome models with linear mixed effects and either Bayesian or EM-based estimation. \cite{tong2023bayesian,wang2024mixed} While these approaches are interpretable and computationally convenient, they rely on linear specifications that may miss nonlinearities and high order interactions in rich baseline covariates, and they may be sensitive to functional-form misspecification.

Another common complication in pragmatic gerontologic trials is missing data. Beyond death truncation, loss to follow-up is frequent as participants relocate, withdraw consent, or become unable to complete questionnaires due to cognitive or physical decline. \cite{little2012prevention,national2011prevention} In such settings, outcomes may be missing even among survivors, creating additional uncertainty in treatment effect estimation. Moreover, survival status itself can sometimes be unobserved—for example, when participants move outside the health system or registry coverage—blurring the distinction between death-related truncation and standard missingness. \cite{chatfield2005systematic} These challenges are especially pronounced in aging populations, where comorbidity, frailty, and mobility constraints increase the risk of incomplete follow-up \cite{chatfield2005systematic, agogo2018longitudinal}. For the estimation of treatment effect heterogeneity, failure to account for missing data can distort subgroup contrasts or mask signals of effect modification. Analyses limited to complete cases implicitly assume that attrition is independent of both treatment and prognosis, an assumption rarely plausible in gerontological research. \cite{cao2022review} Consequently, principled approaches that explicitly model missingness mechanisms under reasonable data assumptions are necessary to recover credible subgroup inferences. \cite{little2012prevention,seaman2013review}

This paper develops an integrated machine learning model tailored to pragmatic CRTs that integrates mixed-effects BART into both the principal strata model and outcome models, while accounting for missing data in survival status and QoL outcomes. By augmenting BART with cluster-level random intercepts, the method preserves flexibility for nonlinear covariate effects and interactions while respecting within-cluster dependence \cite{dorie2022stan,spanbauer2021nonparametric,wundervald2022hierarchical}. To accommodate incomplete follow-up common in pragmatic gerontologic trials, we adopt a nested missing at random scheme that treats missing survival status and, conditional on observed survival, missing outcomes as ignorable given treatment assignment and covariates \cite{tong2025bayesian}. Within this framework, we estimate SACE and CSACE and incorporate an exploratory, post-hoc “fit-the-fit” step that regresses individual-level CSACE estimates among likely always-survivors on baseline covariates to visualize effect modification. Our empirical application revisits the Whole Systems Demonstrator (WSD) telecare trial, focusing on EQ-5D–VAS at 12 months, and illustrates how conditional survivor causal effects can differ from the marginal intent-to-treat effect. The remainder of the paper presents the design and assumptions, the principal stratification estimands and missing data framework, the likelihood and mixed-effects BART specification, a simulation study confirming the performance of our model and algorithm, the WSD analysis and exploratory heterogeneity results, and concluding remarks.

\section{Whole Systems Demonstrator Telecare Questionnaire Study}
\label{sec:data-wsd}

Whole Systems Demonstrator (WSD) Telecare Questionnaire Study \cite{henderson2013cost,hirani2014effect} is a cluster-randomized trial conducted in England during 2008–2009. Randomization occurred at the general practice level, where a total of 1,189 participants across 204 practices in three local authorities (cluster sizes 1–26) were randomized to the telecare intervention or control with roughly a 1:1 ratio. The telecare intervention comprised in-home electronic sensors providing safety monitoring (e.g., fall and hazard alerts), while the control arm received usual care. Our primary analysis focuses on the QoL measured at 12 months by the EQ-5D–VAS index (range 0–100; higher values indicating a better quality of life). \cite{feng2014assessing} Baseline characteristics including age, ethnicity, sex, highest level of education, number of comorbid conditions, living alone status, baseline deprivation score, physical health score, mental health score and EQ-5D–VAS index score by arm are summarized in Table~\ref{tab:baseline}. Baseline deprivation was measured with the English Index of Multiple Deprivation (IMD 2007), linked from participant postcodes to their Lower-layer Super Output Area; the IMD combines seven domains—income, employment, health and disability, education, crime, barriers to housing and services, and living environment—into an area-level score in which higher values indicate greater deprivation. \cite{noble2008english} In WSD, IMD scores ranged from 1.89 to 66.49. Overall, characteristics were broadly comparable between arms: means of the continuous measures and the distributions of sex and ethnicity were closely aligned (all $p\ge 0.31$). The only notable difference was a modest shift in education levels (Cochran--Armitage $p=0.014$), with the intervention arm showing slightly higher proportions at the university/graduate categories.

\begin{table}[htbp]
\centering
\begin{threeparttable}
\caption{Baseline covariates in the Whole Systems Demonstrator (WSD) Telecare Questionnaire Study}
\label{tab:baseline}
\begin{tabular}{lcccc}
\toprule
\textbf{Covariate} & \textbf{Total} & \textbf{Intervention} & \textbf{Control} & \textbf{$p$-value} \\
\midrule
Age, years & 74.13 (13.94) & 73.92 (14.32) & 74.31 (13.62) & 0.634 \\
Deprivation score & 28.13 (15.04) & 27.70 (14.35) & 28.49 (15.61) & 0.363 \\
No. of comorbid conditions & 1.09 (1.45) & 1.07 (1.47) & 1.11 (1.44) & 0.659 \\
Physical health score & 28.09 (8.62) & 28.30 (8.71) & 27.91 (8.55) & 0.436 \\
Mental health score & 33.05 (7.87) & 33.00 (7.98) & 33.10 (7.78) & 0.833 \\
EQ-5D-VAS index score & 53.04 (22.02) & 52.77 (22.07) & 53.27 (22.00) & 0.697 \\
\addlinespace
\textit{Ethnicity} & & & & 0.771 \\
\quad White & 1053 (88.56\%) & 485 (88.18\%) & 568 (88.89\%) & \\
\quad Non-white & 136 (11.44\%) & 65 (11.82\%) & 71 (11.11\%) & \\
\addlinespace
\textit{Sex} & & & & 0.310 \\
\quad Male & 765 (64.34\%) & 345 (62.73\%) & 420 (65.73\%) & \\
\quad Female & 424 (35.66\%) & 205 (37.27\%) & 219 (34.27\%) & \\
\addlinespace
\textit{Highest level of education} & & & & 0.014 \\
\quad No formal education & 779 (65.52\%) & 357 (64.91\%) & 422 (66.04\%) & \\
\quad GCSE/O-levels & 224 (18.84\%) & 92 (16.73\%) & 132 (20.66\%) & \\
\quad A-levels/HNC & 72 (6.06\%) & 30 (5.45\%) & 42 (6.57\%) & \\
\quad University level & 41 (3.45\%) & 26 (4.73\%) & 15 (2.35\%) & \\
\quad Graduate or professional & 73 (6.14\%) & 45 (8.18\%) & 28 (4.38\%) & \\
\addlinespace
\textit{Living alone} & & & & 0.603 \\
\quad Yes & 629 (52.90\%) & 286 (52.00\%) & 343 (53.68\%) & \\
\quad No  & 560 (47.10\%) & 264 (48.00\%) & 296 (46.32\%) & \\
\bottomrule
\end{tabular}
\begin{tablenotes}[flushleft]
\footnotesize
\item Values are mean (SD) for continuous variables and $n$ (\%) for categorical variables. 
$p$-values are from one-way ANOVA for continuous variables and Pearson’s $\chi^2$ test (Fisher’s exact test when expected cell counts $<5$) for categorical variables; for ordered education levels, the $p$-value is from the Cochran--Armitage trend test.
\end{tablenotes}
\end{threeparttable}
\end{table}

For the purposes of principal stratification, we classify a participant as a \emph{nonsurvivor} when the 12 month status is either death or being too ill to respond (due to dementia or loss of capacity, transition to long-term residential/nursing care, including sheltered housing, or data collection was fully reliant on a family caregiver). In the assembled data, 127 participants (10.7\%) had outcomes truncated by death or serious deterioration; 62 (5.2\%) were presumed alive but had unobserved outcomes for non–health-related reasons; for example, some moved out of the area to a non-participating general practice, and others no longer wished to remain in either the intervention or control group; and 237 (19.9\%) had both survival status and the final outcome unobserved (e.g., prolonged hospitalization, refusal to share data, or equipment malfunction).

\section{Methods}
\label{sec:3}

\subsection{Notation and setup}

We study the two-arm cluster-randomized WSD trial with clusters $i=1,\dots,n$,
where cluster $i$ contains $N_i$ individuals $j=1,\dots,N_i$. Let $Z_i\in\{0,1\}$ denote the cluster-level assignment ($Z_i=1$ treatment; $Z_i=0$ control). For each individual $(i,j)$ and $z\in\{0,1\}$, $S_{ij}(z)\in\{0,1\}$ is the potential survival status at end of follow-up ($1$=survive, $0$=death) and $Y_{ij}(z)$ is the corresponding potential non-mortality outcome, which is only well-defined when $S_{ij}(z)=1$; when $S_{ij}(z)=0$ the outcome is undefined and denoted by $\star$ \cite{zhang2003estimation}. The observed data are $(Z_i, S_{ij}, Y_{ij}^{\text{obs}})$ with the usual consistency: $S_{ij}=S_{ij}(Z_i)$ and, if $S_{ij}=1$, then $Y_{ij}^{\text{obs}}=Y_{ij}(Z_i)$; otherwise $Y_{ij}^{\text{obs}}=\star$.

We adopt two assumptions to connect the cluster-level assignment to counterfactual outcomes. 
First, we invoke the Stable Unit Treatment Value Assumption (SUTVA) with partial interference: for each individual $(i,j)$, the potential survival and outcome $\{S_{ij}(z),Y_{ij}(z)\}$ depend only on the assignment of their own cluster $Z_i$; treatment versions are unique and there is no interference across clusters, while arbitrary dependence among individuals within the same cluster is allowed. 
Second, we assume cluster-level randomization with positivity: for every cluster $i$, the collection $\{Y_{ij}(1),Y_{ij}(0),S_{ij}(1),S_{ij}(0)\}_{j=1}^{N_i}$ is independent of $Z_i$, and both arms occur with positive probability, $0<\Pr(Z_i=1)<1$. 
Taken together, these conditions render $Z_i$ ignorable for causal inference at the cluster level and underpin identification based on comparisons of observed outcomes across treatment arms.

These conditions are substantively plausible in the WSD setting. Randomization occurred at the general-practice (cluster) level and assignment was well balanced (639 telecare vs.\ 550 usual care), so positivity holds and $Z_i$ is independent of counterfactual outcomes by design. \cite{henderson2013cost,hirani2014effect} The intervention consisted of standardized, home-installed telecare sensors delivered only to practices assigned to the telecare arm, whereas control practices continued usual care; this delivery model limits cross-practice contamination. Any dependence among individuals within a practice (e.g., shared clinical workflows or referral patterns) is accommodated by allowing arbitrary within-cluster correlation in our models, while interference across practices is unlikely because equipment installation and monitoring are practice-specific and home-based. Empirically, baseline characteristics were similar across arms (Table~\ref{tab:baseline}), consistent with successful randomization. Taken together, these features support SUTVA with partial interference and cluster-level ignorability for $Z_i$ in the WSD trial.

\subsection{Principal stratification and estimands}

We use the principal stratification framework \cite{frangakis2002principal, wang2024mixed} to classify each individual by the pair of potential survival indicators $\big(S_{ij}(1),S_{ij}(0)\big)$. The four strata are:
\begin{itemize}
    \item \textit{always-survivors} $(1,1)$: survive under both treatment and control;
    \item \textit{protected} $(1,0)$: survive only if treated;
    \item \textit{harmed} $(0,1)$: survive only if assigned to control;
    \item \textit{never-survivors} $(0,0)$: die under both treatment and control.
\end{itemize}
For notational convenience, let $G_{ij}\in\{00,10,01,11\}$ encode principal strata membership, where $G_{ij}=ab$ corresponds to $S_{ij}(1)=a$ and $S_{ij}(0)=b$ (e.g., $G_{ij}=11$ for always-survivors, $G_{ij}=10$ for protected, $G_{ij}=01$ for harmed, $G_{ij}=00$ for never-survivors).

We adopt a monotonicity assumption \cite{tong2023bayesian,angrist1996identification} stating that treatment does not reduce survival; equivalently, the harmed stratum is absent and $G_{ij}\in\{11,10,00\}$. Under this condition, observed data identify strata membership for two groups and leave two as mixtures: survivors observed under control must be always-survivors ($G_{ij}=11$), and non-survivors observed under treatment must be never-survivors ($G_{ij}=00$); by contrast, non-survivors under control comprise a mixture of protected and never-survivors ($G_{ij}\in\{10,00\}$), while survivors under treatment comprise a mixture of protected and always-survivors ($G_{ij}\in\{10,11\}$). In the WSD context, where telecare aims to improve well-being, monotonicity is substantively plausible.

Because the pair of non-mortality potential outcomes $\big(Y_{ij}(1),Y_{ij}(0)\big)$ is well-defined only for always-survivors, our primary estimand is the \emph{Survivor Average Causal Effect} (SACE):
\[
\Delta_{\mathrm{SACE}}
=\mathbb{E}\!\left[Y_{ij}(1)-Y_{ij}(0)\,\middle|\,G_{ij}=11\right].
\]
Let $X_{ij}$ denote baseline covariates (individual- and cluster-level). We also consider the \emph{Conditional SACE} (CSACE),
\[
\Delta_{\mathrm{CSACE}}(x)
=\mathbb{E}\!\left[Y_{ij}(1)-Y_{ij}(0)\,\middle|\,X_{ij}=x,\ G_{ij}=11\right],
\]
to characterize heterogeneity among always-survivors. In the WSD trial, we estimate both SACE and CSACE for EQ-5D-VAS.

\subsection{Missing outcome and survival status}
\label{sec:3.3}

In the WSD trial, incompleteness arises from two sources: missing survival status at the end of follow-up and missing non-mortality outcomes (e.g., QoL measures). Let $R_{ij}^{S}$ and $R_{ij}^{Y}$ indicate whether survival status and the final outcome are observed, respectively ($1$ for observed and $0$ for missing). When $S_{ij}=0$, the non-mortality outcome is undefined due to truncation by death, denoted $Y_{ij}^{\text{obs}}=\star$. For reference, we summarize the feasible observation patterns and the recorded variables $\mathcal{O}_{ij}$ in Table~\ref{tab:missing-patterns}; we will refer to each pattern by its label in bold.

\begin{table}[H]
\centering
\caption{Observation patterns and recorded data for individual $(i,j)$.}
\label{tab:missing-patterns}
\renewcommand{\arraystretch}{1.1}
\setlength{\tabcolsep}{6pt}
\begin{tabular}{p{0.43\linewidth} p{0.52\linewidth}}
\hline
\textbf{Pattern (label)} & \textbf{Recorded data} $\mathcal{O}_{ij}$\\
\hline
\textbf{Complete survivor} & $\{Z_i,\ X_{ij},\ S_{ij}=1,\ R_{ij}^S=1,\ R_{ij}^Y=1,\ Y_{ij}^{\text{obs}}\}$\\
\textbf{Death truncation} & $\{Z_i,\ X_{ij},\ S_{ij}=0,\ R_{ij}^S=1\}$ \ (outcome undefined, $Y_{ij}^{\text{obs}}=\star$)\\
\textbf{Survivor, outcome missing} & $\{Z_i,\ X_{ij},\ S_{ij}=1,\ R_{ij}^S=1,\ R_{ij}^Y=0\}$\\
\textbf{Status and outcome missing} & $\{Z_i,\ X_{ij},\ R_{ij}^S=0\}$\\
\hline
\end{tabular}
\end{table}

We formalize missingness using a nested MAR specification. Let $X_{ij}$ collect individual- and cluster-level covariates:
\begin{align}
\text{(A1)}\quad & R_{ij}^S \;\perp\; \big(S_{ij}(1),S_{ij}(0),Y_{ij}(1),Y_{ij}(0)\big)
\;\big|\; (Z_i, X_{ij}), \label{ass:A1}\\[4pt]
\text{(A2)}\quad & R_{ij}^Y \;\perp\; \big(Y_{ij}(1),Y_{ij}(0)\big)
\;\big|\; (Z_i, X_{ij},\ S_{ij}=1,\ R_{ij}^S=1). \label{ass:A2}
\end{align}

Assumption \eqref{ass:A1} states that, conditional on $(Z_i,X_{ij})$, whether survival is recorded does not depend on counterfactual survivals or outcomes; Assumption \eqref{ass:A2} restricts outcome missingness to be MAR within the observed survivor subset ($S_{ij}=1$, $R_{ij}^S=1$). These conditions are compatible with cluster-level randomization and allow arbitrary within-cluster dependence, rendering the missingness mechanism ignorable for likelihood- or Bayesian-based inference when coupled with appropriate models for survival and outcomes.


In the WSD trial, Assumptions~\eqref{ass:A1}-\eqref{ass:A2} encode a \emph{nested} MAR structure: survival status is first subject to missingness through $R_{ij}^S$, and, conditional on being classified as a survivor with observed status ($S_{ij}=1$, $R_{ij}^S=1$), the final outcome is then subject to additional missingness through $R_{ij}^Y$. As described in Section~\ref{sec:data-wsd}, incomplete follow-up in WSD arose from three broad mechanisms: (i) truncation by death or serious clinical deterioration, which we treat as defining the nonsurvivor category for principal stratification; (ii) participants presumed alive but with unobserved QoL outcomes for non–health-related reasons (e.g., relocation or withdrawal from the study); and (iii) cases in which both survival status and QoL were unobserved due to administrative or logistical issues such as prolonged hospitalization, refusal to share records, or equipment problems. 

Under cluster-level randomization, it is plausible that, conditional on the randomized treatment assignment $Z_i$ and the rich baseline covariates $X_{ij}$, the probability that survival status is recorded is driven primarily by these observed prognostic and administrative factors rather than by the unobserved potential survival or QoL outcomes themselves, which is exactly the content of Assumption~\eqref{ass:A1}. Likewise, among individuals with observed survival ($S_{ij}=1$, $R_{ij}^S=1$), missingness in the EQ-5D-VAS outcome is mainly attributable to patterns of loss to follow-up, non-response, or study withdrawal that are well captured by $(Z_i,X_{ij})$, motivating Assumption~\eqref{ass:A2} that $R_{ij}^Y$ is independent of the unobserved potential QoL outcomes given these observed quantities. In this sense, the nested MAR assumption provides a study-specific and practically interpretable approximation to the missing-data mechanism in the presence of truncation by death. To assess the impact of potential deviations from nested MAR, we further conduct simulation-based sensitivity analyses in which the missingness indicators are allowed to depend on latent prognostic variables beyond $(Z_i,X_{ij})$ (Section~\ref{sec:sim-sensitivity}).


In our analysis, we treat missing survival status and missing EQ-5D-VAS values as additional unknown quantities and impute them under the nested MAR assumptions in \eqref{ass:A1}-\eqref{ass:A2}. The imputation scheme closely follows the four patterns in Table~\ref{tab:missing-patterns} and is embedded within our Bayesian estimation as a data–augmentation step.

First, for individuals in the ``Survivor, outcome missing'' pattern ($S_{ij}=1$, $R_{ij}^S=1$, $R_{ij}^Y=0$), we assume that, conditional on $(Z_i,X_{ij})$, the outcome is missing at random. At each iteration of the Gibbs sampler, we draw the missing $Y_{ij}^{\text{mis}}$ from the arm- and stratum-specific outcome model
\[
Y_{ij}^{\text{mis}} \mid (Z_i,X_{ij},S_{ij}=1,G_{ij}=g,\theta_Y) 
\;\sim\; f_Y\!\big(\cdot \mid Z_i,X_{ij},G_{ij}=g,\theta_Y\big),
\]
using the current draws of the model parameters $\theta_Y$ and the principal stratum label $G_{ij}$.

Second, for individuals in the ``Status and outcome missing'' pattern ($R_{ij}^S=0$), both $S_{ij}$ and $Y_{ij}$ are unobserved. For these subjects, we first sample the principal stratum membership $G_{ij}\in\{11,10,00\}$ from the nested Probit membership model described in Section~\ref{sec:3.5}, conditional on $(Z_i,X_{ij})$ and the current latent variables $(Q_{ij},W_{ij})$. Under the monotonicity assumption, $(G_{ij},Z_i)$ then determines survival deterministically: $S_{ij}=1$ if $G_{ij}=11$, or if $G_{ij}=10$ and $Z_i=1$, and $S_{ij}=0$ otherwise. For those with imputed $S_{ij}=1$, we draw $Y_{ij}^{\text{mis}}$ from the corresponding arm- and stratum-specific outcome model as above; for those with imputed $S_{ij}=0$, the non-mortality outcome remains undefined by design ($Y_{ij}^{\text{obs}}=\star$) and is not imputed.

\subsection{Likelihood and Bayesian principal stratification}
\label{sec:3.4}

We adopt a Bayesian principal stratification framework \cite{hirano2000assessing} to infer principal stratum membership and to estimate SACE/CSACE. We can classify participants by observed treatment assignments and survival status at the final assessment:
\begin{itemize}
  \item $O(1,1)=\{(i,j)\mid Z_i=1,\ S_{ij}=1\}$: participants from treated clusters who survived;
  \item $O(1,0)=\{(i,j)\mid Z_i=1,\ S_{ij}=0\}$: participants from treated clusters who did not survive;
  \item $O(0,1)=\{(i,j)\mid Z_i=0,\ S_{ij}=1\}$: participants from control clusters who survived;
  \item $O(0,0)=\{(i,j)\mid Z_i=0,\ S_{ij}=0\}$: participants from control clusters who did not survive.
\end{itemize}
Under monotonicity, $O(1,0)$ are identified as never-survivors ($G_{ij}=00$) and $O(0,1)$ as always-survivors ($G_{ij}=11$); by contrast, $O(1,1)$ mixes $\{11,10\}$ and $O(0,0)$ mixes $\{10,00\}$. We need additional modeling to deal with this mixture. We use $S^{\text{obs}}$ to denote the observed survival status for all individuals and $Y^{\text{obs}}$ to denote the collection of observed non-mortality outcomes for all individuals.

Let $p_{ij,g}=\Pr(G_{ij}=g\mid X_{ij},\theta_S)$ denote the stratum membership probabilities (summing to one across $g$). For the outcome model, define the arm- and stratum-specific density evaluated at the observed outcome,
\[
f_{ij,g,z}=f_Y\!\big(Y_{ij}^{\text{obs}}\mid G_{ij}=g,\ Z_i=z,\ X_{ij},\ \theta_Y\big),
\]
which is needed only when the outcome is well-defined in stratum $g$ under arm $z$ (i.e., $g=11$ for $z=0$; $g\in\{11,10\}$ for $z=1$). Let $\theta=(\theta_S,\theta_Y)$ and take independent priors $P(\theta)=P(\theta_S)P(\theta_Y)$.

The observed data likelihood (conditional on covariates) is
\[
\begin{aligned}
L(Y^{\text{obs}},S^{\text{obs}},X,Z\mid \theta)
&=\prod_{(i,j)\in O(1,1)}
\Big\{\, p_{ij,11}\, f_{ij,11,1} \;+\; p_{ij,10}\, f_{ij,10,1}\,\Big\} \\
&\quad\times \prod_{(i,j)\in O(1,0)} p_{ij,00} \\
&\quad\times \prod_{(i,j)\in O(0,1)} p_{ij,11}\, f_{ij,11,0} \\
&\quad\times \prod_{(i,j)\in O(0,0)} \big( p_{ij,10} + p_{ij,00} \big).
\end{aligned}
\]
This likelihood highlights where mixtures arise (exactly in $O(1,1)$ and $O(0,0)$); no outcome density is introduced for never-survivors because the non-mortality outcome is undefined. Cluster-level randomization is accommodated by conditioning on $Z_i$, while within-cluster dependence among individuals remains unrestricted.

\subsection{Model specification}
\label{sec:3.5}

We use a nested Probit construction to impute the unobserved principal strata labels $G_{ij}\in\{00,10,11\}$. \cite{chen2024bayesian} 
Two latent Gaussian variables are augmented for each individual: $Q_{ij}$ first separates never-survivors $(00)$ from the survivor capable strata, and $W_{ij}$ then splits the survivor capable group into protected $(10)$ versus always-survivors $(11)$.

\[
\begin{aligned}
& Q_{ij}\mid m_Q(\bullet),\,\mathbf X_{Q,ij}\ \sim\ \mathcal N\!\big(m_Q(\mathbf X_{Q,ij}),\,1\big), 
\quad 
\begin{cases}
G_{ij}=00, & \text{if } Q_{ij}\le 0,\\
G_{ij}\in\{10,11\}, & \text{if } Q_{ij}>0,
\end{cases} \\[1.0ex]
& W_{ij}\mid m_W(\bullet),\,\mathbf X_{W,ij}\ \sim\ \mathcal N\!\big(m_W(\mathbf X_{W,ij}),\,1\big), 
\quad 
\begin{cases}
G_{ij}=10, & \text{if } Q_{ij}>0\ \text{and } W_{ij}\le 0,\\
G_{ij}=11, & \text{if } Q_{ij}>0\ \text{and } W_{ij}> 0.
\end{cases}
\end{aligned}
\]
Here $m_Q(\bullet)$ and $m_W(\bullet)$ are conditional mean functions (e.g., Probit regressions) of covariates $\mathbf X_{Q,ij}$ and $\mathbf X_{W,ij}$; they can include both individual- and cluster-level covariates. Conditional on covariates, we take $Q_{ij}$ and $W_{ij}$ as independent. This yields the strata-membership probabilities
\[
\begin{aligned}
p_{ij,00}&=P(G_{ij}=00\mid \mathbf X_{Q,ij})
= 1-\Phi\!\big(m_Q(\mathbf X_{Q,ij})\big), \\[0.8ex]
p_{ij,10}&=P(G_{ij}=10\mid \mathbf X_{Q,ij},\mathbf X_{W,ij})
= \Phi\!\big(m_Q(\mathbf X_{Q,ij})\big)\,\big\{1-\Phi\!\big(m_W(\mathbf X_{W,ij})\big)\big\}, \\[0.8ex]
p_{ij,11}&=P(G_{ij}=11\mid \mathbf X_{Q,ij},\mathbf X_{W,ij})
= \Phi\!\big(m_Q(\mathbf X_{Q,ij})\big)\,\Phi\!\big(m_W(\mathbf X_{W,ij})\big),
\end{aligned}
\]
where $\Phi(\cdot)$ is the standard normal CDF.

For outcomes, we specify arm- and stratum-specific Gaussian models (defined only when the potential outcome is meaningful in that stratum/arm):
\[
\{\,Y_{ij}(z)\mid G_{ij}=g,\ m_{g,z}(\bullet),\ \mathbf X_{g,z,ij}\,\}
\ \sim\ \mathcal N\!\big(m_{g,z}(\mathbf X_{g,z,ij}),\ \sigma^2_{g,z}\big),
\]
with $z\in\{0,1\}$ when $g=11$ and $z=1$ when $g=10$. Concretely,
\[
\begin{aligned}
Y_{ij}(1)\mid (G_{ij}=11) &\sim \mathcal N\!\big(m_{11,1}(\mathbf X_{11,1,ij}),\,\sigma^2_{11,1}\big),\\
Y_{ij}(0)\mid (G_{ij}=11) &\sim \mathcal N\!\big(m_{11,0}(\mathbf X_{11,0,ij}),\,\sigma^2_{11,0}\big),\\
Y_{ij}(1)\mid (G_{ij}=10) &\sim \mathcal N\!\big(m_{10,1}(\mathbf X_{10,1,ij}),\,\sigma^2_{10,1}\big),
\end{aligned}
\]
while $Y_{ij}(0)$ is undefined for $G_{ij}=10$ and both $Y_{ij}(0),Y_{ij}(1)$ are undefined for $G_{ij}=00$ by truncation-by-death. Completing the specification amounts to choosing the systematic components for the latent–Probit and outcome models, namely 
$m_Q(\mathbf X_{Q,ij})$, $m_W(\mathbf X_{W,ij})$, and $m_{g,z}(\mathbf X_{g,z,ij})$. 
A conventional choice is to use (generalized) linear mixed models with cluster-level random effects, which is convenient and interpretable. 
However, such models impose linear and additive effects (and require pre-specifying interactions), an assumption that can be too restrictive in CRT settings with potential nonlinearity and treatment–covariate interactions; misspecification may bias strata probabilities and, in turn, SACE/CSACE. 
Motivated by this, in what follows we adopt more flexible specifications for these mean functions—preserving random effects to account for clustering while allowing nonlinearities and high order interactions.

\subsection{Integrating mixed-effects BART into principal stratification}

To flexibly represent nonlinearities and higher order interactions in the conditional mean functions, we use Bayesian Additive Regression Trees (BART) \cite{chipman2010bart}, which approximates a regression surface by a sum of shallow trees,
\begin{equation}\label{eq:bart}
m(\mathbf X_{ij}) \;=\; \sum_{k=1}^K g_k\!\big(\mathbf X_{ij};\,\mathcal J_k,\mathcal M_k\big),
\end{equation}
where each \(g_k\) is a regression tree with splitting structure \(\mathcal J_k\) and terminal node parameters \(\mathcal M_k\).
Because cluster-randomized trials exhibit within-cluster correlation, we augment each mean function with a cluster-level random intercept to absorb unobserved heterogeneity while learning covariate effects nonparametrically.

The mixed-effects BART specifications are
\begin{equation*}
m_Q(\mathbf X_{Q,ij})
=\sum_{k=1}^K g_{Q,k}\!\big(\mathbf X_{Q,ij};\,\mathcal J_{Q,k},\mathcal M_{Q,k}\big)+ b_i^{(Q)},
\qquad b_i^{(Q)}\sim \mathcal N(0,\sigma_{b,Q}^2),
\end{equation*}
\begin{equation*}
m_W(\mathbf X_{W,ij})
=\sum_{k=1}^K g_{W,k}\!\big(\mathbf X_{W,ij};\,\mathcal J_{W,k},\mathcal M_{W,k}\big)+ b_i^{(W)},
\qquad b_i^{(W)}\sim \mathcal N(0,\sigma_{b,W}^2),
\end{equation*}

\begin{equation*}
m_{g,z}(\mathbf X_{g,z,ij})
= \sum_{k=1}^K g_{g,z,k}\!\big(\mathbf X_{g,z,ij}; \mathcal J_{g,z,k}, \mathcal M_{g,z,k}\big) + b_i^{(g)},
\qquad (g,z)\in\{(11,0),(11,1),(10,1)\}.
\end{equation*}
\begin{equation*}
b_i^{(g)} \sim \mathcal N(0,\sigma_{b,g}^2).
\end{equation*}

In practice, \(m_Q\) and \(m_W\) are learned by regressing the current draws of the latent membership variables on covariates using Gaussian mixed-effects BART, and \(m_{11,0}\), \(m_{11,1}\), \(m_{10,1}\) are fit on the corresponding arm–stratum subsets; because \(G_{ij}\) is latent, these subsets and the associated means are updated within the MCMC.

Mixed-effects BART has several implementations, including \texttt{stan4bart} \cite{dorie2022stan}, \texttt{mxbart} \cite{spanbauer2021nonparametric}, and \texttt{hebart} \cite{wundervald2022hierarchical}. In simulations aligned with our design, \texttt{mxbart} provided a favorable balance of accuracy and computational efficiency, and we therefore use it to estimate \(m_Q\), \(m_W\), and \(m_{g,z}\). The mxbart model is a general framework for incorporating hierarchical random effects into BART, originally developed for longitudinal clinical trials. This approach supports both random intercepts and slopes, though in CRTs we adopt the special case of a cluster-level intercept. 

\subsection{Model estimation}
A fully Bayesian joint model integrating imputation within the principal stratification sampler is also feasible but beyond the scope of this illustration. Missing baseline covariates were completed using multiple imputation \cite{rubin1978multiple} to obtain a single analysis dataset. The covariate set used in the strata membership and outcome models comprised sex, age, ethnicity, highest education level, an indicator for living alone status, number of comorbid conditions, index of multiple deprivation (IMD) score, physical and mental health scores, and the baseline EQ-5D-VAS index.\cite{hirani2014effect}

We initialize the Markov chain by assigning principal–stratum labels using information identified by monotonicity: treated non-survivors are set to the never-survivor stratum and control survivors to the always-survivor stratum. For individuals in the remaining observed mixtures (treated survivors; control non-survivors), initial labels are drawn at random from the admissible strata given $(Z_i,S_{ij})$. For units with unobserved survival, we impute their initial survival status by BART.  Given these initial labels, initial estimates of outcome model $m_{g,z}(\cdot)$ can be obtained by fitting BART and initial estimate for $\sigma^2_{g,z}$ can be simultaneously obtained from this initial fit. For the strata–membership means $m_Q(\cdot)$ and $m_W(\cdot)$, we obtain starting values from parametric logistic regressions of the initial labels on covariates; latent normals $Q_{ij}$ and $W_{ij}$ are then drawn from the appropriate truncated Gaussian distributions implied by the initial labels. Hyperparameters for $\sigma_{g,z}^2$ use weakly informative inverse-gamma priors with shape and rate $a_0=b_0=0.001$ \cite{chen2024bayesian}. For random intercept $\sigma_{b}^2$, we use the default prior in mxbart: hierarchical Inverse-Wishart with degree of freedom 3.

Each iteration updates: (i) the outcome models $m_{11,1}$, $m_{11,0}$, and $m_{10,1}$ via mixed-effects BART on the current arm–stratum subsets; (ii) the residual variances $\{\sigma^2_{g,z}\}$ by conjugate Gaussian updates; (iii) the strata–membership means $m_Q$ and $m_W$ via mixed-effects BART regressions of the current latent normals $Q_{ij}$ and $W_{ij}$ on covariates; (iv) the latent $Q_{ij}$ and $W_{ij}$ from their truncated normal full conditionals; (v) principal stratum labels $G_{ij}$ using their posterior probabilities that combine the current membership model and outcome likelihood; and (vi) missing data under the nested-MAR structure: survival is sampled from the strata model, outcomes are drawn from the appropriate arm– and stratum–specific outcome model when survival is (imputed) $1$, and outcomes are undefined by design when survival is $0$. We used the default settings of \texttt{mxbart} for Gaussian responses with a random intercept per cluster. 

We ran 20{,}000 iterations, discarding the first 10{,}000 as burn-in. Point estimates and 95\% credible intervals for SACE and CSACE are based on posterior draws from the retained iterations. All analyses were conducted in \textsf{R}~4.4.1 using \texttt{mxbart} and \texttt{dbarts}.

\section{Simulation Studies}

\subsection{Benchmark Parametric Mixed-Effects Model}

To provide a parametric benchmark for the proposed mixed-effects BART principal stratification model, we first consider a fully parametric linear mixed-effects specification obtained by replacing all nonparametric mean functions with linear predictors while retaining the same hierarchical structure. Within this framework, the principal-strata membership model is still formulated through the nested Probit construction, but the latent means are modeled linearly with cluster-level random intercepts:
\begin{equation*}
m_Q(\mathbf X_{Q,ij})
= \mathbf X_{Q,ij}^\top \beta_Q + b_i^{(Q)},
\qquad b_i^{(Q)} \sim \mathcal N\!\big(0,\sigma_{b,Q}^2\big),
\end{equation*}
\begin{equation*}
m_W(\mathbf X_{W,ij})
= \mathbf X_{W,ij}^\top \beta_W + b_i^{(W)},
\qquad b_i^{(W)} \sim \mathcal N\!\big(0,\sigma_{b,W}^2\big),
\end{equation*}
with latent Gaussian variables
\[
Q_{ij} = m_Q(\mathbf X_{Q,ij}) + \varepsilon_{Q,ij}, 
\quad 
W_{ij} = m_W(\mathbf X_{W,ij}) + \varepsilon_{W,ij}, 
\quad
\varepsilon_{Q,ij},\varepsilon_{W,ij} \overset{\text{ind}}{\sim} \mathcal N(0,1),
\]
and the same truncation rules as in Section~\ref{sec:3} for determining the principal stratum label \(G_{ij}\in\{00,10,11\}\).

For the outcome model, we specify arm- and stratum-specific Gaussian linear mixed models, with cluster-level random intercepts:
\begin{equation*}
m_{g,z}(\mathbf X_{g,z,ij})
= \mathbf X_{g,z,ij}^\top \alpha_{g,z} + b_i^{(g)},
\qquad (g,z)\in\{(11,0),(11,1),(10,1)\},
\end{equation*}
\begin{equation*}
b_i^{(g)} \sim \mathcal N\!\big(0,\sigma_{b,g}^2\big),
\qquad
Y_{ij}(z)\mid G_{ij}=g \sim \mathcal N\!\big(m_{g,z}(\mathbf X_{g,z,ij}),\,\sigma_{g,z}^2\big),
\end{equation*}
where, as before, \(Y_{ij}(z)\) is only defined when the potential outcome is meaningful in stratum \(g\) under arm \(z\) (i.e., \(g=11\) for \(z\in\{0,1\}\) and \(g=10\) for \(z=1\)).

For posterior inference under this parametric specification, we develop a closed-form Gibbs sampler in which all full conditional distributions are available in standard conjugate form. We place independent Gaussian priors on the fixed-effects coefficients and inverse-gamma priors on the variance components:
\begin{align*}
&\beta_Q \sim \mathcal N\!\big(0,\Lambda_Q^{-1}\big), 
\qquad
\beta_W \sim \mathcal N\!\big(0,\Lambda_W^{-1}\big),\\
&\alpha_{g,z} \sim \mathcal N\!\big(0,\Lambda_{g,z}^{-1}\big), 
\qquad (g,z)\in\{(11,0),(11,1),(10,1)\},\\[2pt]
&\sigma_{b,Q}^2 \sim \text{IG}\big(a_{b,Q},b_{b,Q}\big), 
\qquad
\sigma_{b,W}^2 \sim \text{IG}\big(a_{b,W},b_{b,W}\big),\\
&\sigma_{b,g}^2 \sim \text{IG}\big(a_{b,Y},b_{b,Y}\big),
\qquad
\sigma_{g,z}^2 \sim \text{IG}\big(a_{\sigma,g,z},b_{\sigma,g,z}\big),
\quad (g,z)\in\{(11,0),(11,1),(10,1)\},
\end{align*}
with weakly informative hyperparameters (e.g., all inverse-gamma priors use \(a=b=0.001\)).

Given these priors, the Gibbs sampler alternates between updating:
(i) the outcome regression coefficients \(\{\alpha_{g,z}\}\) and random intercepts \(\{b_i^{(g)}\}\) from multivariate Gaussian full conditionals;
(ii) the outcome variances \(\{\sigma_{g,z}^2\}\) and random-intercept variances \(\{\sigma_{b, g}^2\}\) from inverse-gamma full conditionals;
(iii) the strata-membership coefficients \(\beta_Q,\beta_W\) and their random intercepts \(b_i^{(Q)},b_i^{(W)}\) using Gaussian updates;
(iv) the latent normals \(Q_{ij},W_{ij}\) from their truncated normal full conditionals implied by the current \(G_{ij}\); and
(v) the principal stratum labels \(G_{ij}\) and any missing survival or outcome data, using the same nested Probit structure and likelihood factorization as in Section~\ref{sec:3.4} under the nested MAR assumptions. 

This parametric linear mixed-effects principal stratification model thus provides a natural ``LMM'' analogue of our proposed mixed-effects BART approach, and serves as a useful comparator in the simulation studies below. In our simulation study, we consider four different approaches to estimate the CSACE and principal stratum membership: (i) the proposed mixed-effects Bayesian machine learning approach (YBSB), in which all mean functions in both the outcome and stratum models are specified nonparametrically using mixed-effects BART as implemented in \texttt{mxbart}; (ii) the YBSP approach, where the mean functions in the outcome models are specified nonparametrically using mixed-effects BART, while those in the stratum model are specified using linear mixed-effects parametric models; (iii) the YPSB approach, where the mean functions in the outcome models are specified using linear mixed-effects parametric models, while those in the stratum model are specified nonparametrically using mixed-effects BART; and finally (iv) the fully parametric approach (YPSP), in which all mean functions in both the outcome and stratum models are specified using linear mixed-effects parametric models.

\subsection{Simulation Design}

We generated clustered data under a design closely mirroring the WSD setting but with known ground truth. Each simulated dataset consists of \(I=200\) clusters, each containing \(N_i = 6\) individuals, yielding a total sample size of \(N = 1200\). Clusters were randomized to treatment or control using balanced complete randomization with a 1:1 allocation, so that exactly half of the clusters received treatment (\(Z_i=1\)) and the remainder received control (\(Z_i=0\)). Each individual inherits the cluster assignment \(Z_i\). We generated five individual-level baseline variables \(\mathbf X_{ij} = (X_{ij1},\dots,X_{ij5})\) with
\(
X_{ij1}, X_{ij2}, X_{ij3} \;\sim\; \mathcal N(0,1)\) and
\(
X_{ij4}, X_{ij5} \;\sim\; \text{Bernoulli}(0.5),
\) 
independently across individuals. These covariates enter both the principal-strata membership model and the outcome model through nonlinear and interaction terms.

Principal strata were generated using the nested Probit construction described in Section~\ref{sec:3.4}. For each individual we first draw cluster-level random intercepts
\(
b_i^{(Q)}\overset{\text{i.i.d.}}{\sim}\mathcal N(0,\sigma_{b,Q}^2),\quad b_i^{(W)}\overset{\text{i.i.d.}}{\sim}\mathcal N(0,\sigma_{b,W}^2),
\quad \sigma_{b,Q}^2 = \sigma_{b,W}^2 = 0.0204,
\) 
and individual-level errors
\(
\varepsilon_{Q,ij} \;\overset{\text{i.i.d.}}{\sim}\; \mathcal N(0,1), \quad \varepsilon_{W,ij} \overset{\text{i.i.d.}}{\sim}\; \mathcal N(0,1)
\) 
and define the latent means
\[
m_Q(\mathbf X_{ij}) 
= 1.2
+ 0.6 X_{ij1}
- 0.7 \cos(X_{ij2})
+ 1.5 X_{ij3}^2
+ 0.7 X_{ij1} X_{ij2}
- 0.5 \sin\!\big(\pi X_{ij1} X_{ij3}\big)
- 0.3 X_{ij4}
+ 1.0 X_{ij5}
+ b_i^{(Q)},
\]
\[
m_W(\mathbf X_{ij}) 
= 0.9
+ 0.9\, \operatorname{expit}\!\big(1.5 X_{ij1}\big)
- 0.6 X_{ij2}
+ 0.4 X_{ij3}
+ 0.5 X_{ij1} X_{ij3}
- 0.4 \cos\!\big(\pi X_{ij2}\big)
- 0.3 X_{ij4}
+ 0.9 X_{ij5}
+ b_i^{(W)}.
\]

The latent Gaussian variables are then
\[
Q_{ij} = m_Q(\mathbf X_{ij}) + \varepsilon_{Q,ij}, 
\qquad
W_{ij} = m_W(\mathbf X_{ij}) + \varepsilon_{W,ij},
\]
and principal stratum membership is determined by the truncation rule
\[
G_{ij} =
\begin{cases}
00, & Q_{ij} \le 0,\\[2pt]
10, & Q_{ij} > 0,\ W_{ij} \le 0,\\[2pt]
11, & Q_{ij} > 0,\ W_{ij} > 0,
\end{cases}
\]
corresponding to never-survivors \((00)\), protected \((10)\), and always-survivors \((11)\), respectively, under the monotonicity assumption. We also keep a numeric coding \(G_{ij}^{\text{num}}\in\{0,1,2\}\) for convenience in computation.

Given \(G_{ij}\), potential survival indicators \(\big(S_{ij}(0), S_{ij}(1)\big)\) are assigned deterministically in accordance with the principal stratification.
The realized survival status under the randomized arm is
\(
S_{ij}^{\text{true}} = S_{ij}(Z_i).
\)

Potential non-mortality outcomes are generated only when they are well-defined. We introduce a cluster-level random effect for the outcome,
\(
b_i^{(Y)} \sim \mathcal N\!\big(0,\sigma_{b,Y}^2\big),
\sigma_{b,Y}^2 = 0.02,
\)
and individual-level errors
\(
\varepsilon_{ij}^{(Y)} \sim \mathcal N\!\big(0,\sigma_{\varepsilon,Y}^2\big),
\sigma_{\varepsilon,Y}^2 = 0.98,
\)
so that the marginal outcome variance is approximately one and the outcome intraclass correlation is \(\text{ICC}\approx 0.02\), which matches the previous analysis of WSD. \cite{tong2023bayesian}

For always-survivors (\(G_{ij}=11\)), we specify a nonlinear baseline outcome function and heterogeneous treatment effect:
\[
f_{11}(\mathbf X_{ij}) 
= 0.2
+ 0.2 X_{ij1}
- 0.6 \sin(X_{ij2})
+ 0.3 X_{ij1} X_{ij2}
+ 0.2\big(X_{ij3} - 0.5\big)^3
- 0.3 X_{ij5},
\]
\[
\tau_{11}(\mathbf X_{ij}) 
= -0.3
+ 0.5 \operatorname{expit}\!\big(2 X_{ij1}\big)
- 0.7 \cos\!\big(\pi X_{ij2} X_{ij3}\big)
- 0.4 X_{ij4}
+ 0.2 X_{ij5}.
\]

The corresponding potential outcomes are
\[
Y_{ij}(0) = f_{11}(\mathbf X_{ij}) + b_i^{(Y)} + \varepsilon_{ij}^{(Y)}, 
\qquad
Y_{ij}(1) = f_{11}(\mathbf X_{ij}) + \tau_{11}(\mathbf X_{ij}) + b_i^{(Y)} + \varepsilon_{ij}^{(Y)},
\quad \text{for } G_{ij}=11.
\]

For protected individuals (\(G_{ij}=10\)), the potential outcome under control is undefined by truncation-by-death, and we only specify \(Y_{ij}(1)\):
\[
f_{10}(\mathbf X_{ij}) 
= -1.5
+ 0.5 X_{ij1}
+ 0.8 X_{ij2}^2
- 0.6 \arctan(X_{ij3})
- 0.5 X_{ij4},
\]
\[
\tau_{10}(\mathbf X_{ij}) 
= 0.2
+ 0.15 \log\!\big(\lvert 0.5 + X_{ij1}\rvert + 10^{-4}\big)
- 0.1 X_{ij2}^2
+ 0.3 \sin\!\big(\pi X_{ij2} X_{ij3}\big)
- 0.2 X_{ij4}
+ 0.05 X_{ij5}.
\]

\[
Y_{ij}(1) = f_{10}(\mathbf X_{ij}) + \tau_{10}(\mathbf X_{ij}) + b_i^{(Y)} + \varepsilon_{ij}^{(Y)},
\quad \text{for } G_{ij}=10,
\]
while \(Y_{ij}(0)\) is left undefined. For never-survivors (\(G_{ij}=00\)), both \(Y_{ij}(0)\) and \(Y_{ij}(1)\) are undefined.

To emulate the nested MAR mechanism in Section~\ref{sec:3.3}, we then impose additional missingness on survival status and outcomes. First, the survival-status indicator \(R_{ij}^S\) (1 if survival is recorded, 0 if missing) is generated via a logistic regression depending on treatment and covariates:
\[
\Pr\big(R_{ij}^S = 1 \mid Z_i, \mathbf X_{ij}\big)
= \operatorname{expit}\!\big(1.8 + 0.4 Z_i - 0.3 X_{ij1} + 0.5 X_{ij2} - 0.2 X_{ij3}\big),
\]
and the recorded survival is
\[
S_{ij}^{\text{obs}} =
\begin{cases}
S_{ij}^{\text{true}}, & R_{ij}^S = 1,\\
\text{NA},            & R_{ij}^S = 0.
\end{cases}
\]
Whenever \(R_{ij}^S = 0\), the outcome is also set to missing. Among individuals who truly survive and have recorded survival (\(S_{ij}^{\text{true}}=1, R_{ij}^S=1\)), we generate an outcome-response indicator \(R_{ij}^Y\) according to
\[
\Pr\big(R_{ij}^Y = 1 \mid Z_i, \mathbf X_{ij}, S_{ij}^{\text{true}}=1, R_{ij}^S=1\big)
= \operatorname{expit}\!\big(2.5 + 0.5 Z_i - 0.2 X_{ij2} - 0.6 X_{ij3} + 0.3 X_{ij5}\big),
\]
and define the final observed outcome
\[
Y_{ij}^{\text{obs}} =
\begin{cases}
Y_{ij}(Z_i), & R_{ij}^S = 1,\ R_{ij}^Y = 1,\\
\text{NA},                 & R_{ij}^S = 0\ \text{or}\ (S_{ij}^{\text{true}}=1,\ R_{ij}^S=1,\ R_{ij}^Y=0),\\
\star,                     & S_{ij}^{\text{true}}=0,\ R_{ij}^S=1.
\end{cases}
\]

This data generating mechanism induces nontrivial principal stratum mixtures, nonlinear and heterogeneous treatment effects, cluster-level correlation in both strata and outcomes, and a nested MAR missingness structure. It therefore provides a challenging yet controlled setting to compare the proposed mixed-effects BART principal stratification model with its parametric linear mixed-effects analogue.

\subsection{Performance Metrics}
\label{sec:4.3}

We evaluated the performance of the CSACE estimators from the four approaches using four metrics: precision in estimation of heterogeneous effects (PEHE), absolute bias, expected regret and empirical coverage probability. Because the CSACE is defined among always-survivors, we focused on individuals who are likely to belong to the always-survivor principal stratum.

Specifically, for each simulated dataset we defined the set of \emph{likely always-survivors} as those individuals whose posterior probability of belonging to the always-survivor principal stratum satisfies
\(
\Pr\big(G_{ij}=11 \mid \text{data}\big) \ge 0.8.
\)
We encode these individuals by setting the estimated principal stratum label to \(\widehat{G}_{ij}=11\). Let
\[
\mathcal{A} = \big\{(i,j): \widehat{G}_{ij}=11\big\}, 
\qquad
N_G = |\mathcal{A}|
\]
be the set and total number of likely always-survivors. For each \((i,j)\in\mathcal{A}\), let \(\widehat{\tau}_{ij}\) denote the estimated individual CSACE (posterior mean of the individual treatment effect) and \(\tau_{ij}\) the corresponding true effect from the data generating process. We also denote the \(95\%\) credible interval for \(\tau_{ij}\) by \([\widehat{L}_{ij},\widehat{U}_{ij}]\), obtained from the \(2.5\%\) and \(97.5\%\) posterior quantiles.

The \emph{precision in estimation of heterogeneous effects} (PEHE) is defined as the root mean squared error between the estimated and true conditional treatment effects among likely always-survivors:
\[
\text{PEHE}
= \sqrt{\frac{1}{N_G} \sum_{(i,j)\in\mathcal{A}}
\big\{\widehat{\tau}_{ij} - \tau_{ij}\big\}^2 }.
\]

To assess bias in the estimated individual effects, we computed the \emph{absolute bias},
\[
\text{AbsBias}
= \frac{1}{N_G} \sum_{(i,j)\in\mathcal{A}}
\big|\widehat{\tau}_{ij} - \tau_{ij}\big|
.
\]

To evaluate the utility of the estimated effects for individualized treatment decisions, we computed the \emph{expected regret}. For each individual, the oracle optimal treatment rule assigns treatment if and only if \(\tau_{ij}>0\), whereas the data-driven rule assigns treatment if and only if \(\widehat{\tau}_{ij}>0\). The regret for individual \((i,j)\) is defined as \(|\tau_{ij}|\) if the two rules disagree and \(0\) otherwise. The expected regret is then
\[
\text{Regret}
= \frac{1}{N_G} \sum_{(i,j)\in\mathcal{A}}
\big|\tau_{ij}\big|\,
\mathbb{I}\!\left\{\mathrm{sign}\big(\widehat{\tau}_{ij}\big) \neq \mathrm{sign}\big(\tau_{ij}\big)\right\},
\]
where \(\mathbb{I}(\cdot)\) denotes the indicator function.

To quantify uncertainty calibration, we computed the \emph{empirical coverage probability} of the \(95\%\) credible intervals among likely always-survivors,
\[
\text{Coverage}
= \frac{1}{N_G} \sum_{(i,j)\in\mathcal{A}}
\mathbb{I}\!\left\{\tau_{ij} \in [\widehat{L}_{ij},\widehat{U}_{ij}]\right\},
\]
that is, the proportion of individuals for whom the true effect lies within the estimated \(95\%\) credible interval.

In addition to evaluating CSACE estimation, we also assessed how well each method classifies individuals into the always-survivor principal stratum. Let
\(
p_{ij}^{(11)} = \Pr\big(G_{ij}=11 \mid \text{data}\big)
\)
denote the posterior probability that individual $(i,j)$ belongs to the always-survivor stratum. Given a prespecified threshold $q \in (0,1)$ (in our simulations $q=0.8$), we define the estimated always-survivor indicator
\(
\widehat{A}_{ij} = \mathbb{I}\!\big\{p_{ij}^{(11)} \ge q\big\},
\)
and use the true principal stratum label $G_{ij}$ from the data generating process to construct the oracle indicator
\(
A_{ij}^{\text{true}} = \mathbb{I}\!\big\{G_{ij}=11\big\}.
\)

Because $\mathcal A$ is data-adaptive and can vary across methods, we additionally report principal stratum classification power and FDR to contextualize CSACE estimation accuracy. We summarize principal stratum classification performance using two metrics: the \emph{power} (sensitivity) for identifying always-survivors and the \emph{false discovery rate} (FDR) among those classified as always-survivors. The power is defined as
\[
\text{Power}
= \Pr\big(\widehat{A}_{ij}=1 \,\big|\, A_{ij}^{\text{true}}=1\big)
= \frac{\sum_{i,j} \mathbb{I}\!\big(A_{ij}^{\text{true}}=1,\ \widehat{A}_{ij}=1\big)}
       {\sum_{i,j} \mathbb{I}\!\big(A_{ij}^{\text{true}}=1\big)},
\]
i.e., the proportion of true always-survivors that are correctly classified as such. The FDR is defined as
\[
\text{FDR}
= \Pr\big(A_{ij}^{\text{true}}=0 \,\big|\, \widehat{A}_{ij}=1\big)
= \frac{\sum_{i,j} \mathbb{I}\!\big(A_{ij}^{\text{true}}=0,\ \widehat{A}_{ij}=1\big)}
       {\sum_{i,j} \mathbb{I}\!\big(\widehat{A}_{ij}=1\big)},
\]
that is, the proportion of individuals declared to be always-survivors who in fact belong to other principal strata. 

In the simulation implementation, we take $p_{ij}^{(11)}$ as the posterior mean of $\mathbb{I}\{G_{ij}=11\}$ over retained MCMC iterations, and set $q=0.8$ when computing both power and FDR.

\subsection{Simulation Results}

We conducted a Monte Carlo simulation study with 100 replicated datasets. Across 100 simulated datasets, the average true principal stratum composition was approximately 10.7\% never-survivors ($G=00$), 10.6\% protected ($G=10$), and 78.7\% always-survivors ($G=11$). The corresponding observed-data patterns were, on average, 13.5\% with both survival status and outcome missing, 14.3\% with death truncation, 67.7\% complete survivors with observed outcomes, and 4.4\% survivors with missing outcomes. 
Table~\ref{tab:sim-summary} summarizes the performance of the four estimation approaches using the metrics defined in Section~\ref{sec:4.3}. Overall, the proposed mixed-effects Bayesian machine learning approach with BART in both the outcome and stratum models (YBSB) achieves the best or near-best performance across most criteria.

For estimation of heterogeneous CSACE among likely always-survivors, the two approaches that use mixed-effects BART for the outcome model (YBSB and YBSP) substantially outperform those based on linear mixed-effects outcome models (YPSB and YPSP). 
In particular, YBSB and YBSP yield smaller PEHE and absolute bias, and lower expected regret, whereas YPSB and YPSP exhibit noticeably larger errors. 
This pattern is reflected in interval estimation as well: the empirical coverage of the $95\%$ credible intervals is close to nominal for YBSB and YBSP (0.951–0.959), but markedly below nominal (around 0.45) for YPSB and YPSP, indicating that misspecification of the outcome model leads to severe undercoverage even when the principal stratum model is correctly specified. 
These findings are consistent with previous work showing that, for heterogeneous treatment effect estimation, flexible modeling of the outcome regressions is particularly important.\cite{chen2024bayesian}

The choice of model for the principal stratum membership mainly affects classification performance. 
When the stratum model is specified nonparametrically via mixed-effects BART (YBSB and YPSB), the power to identify always-survivors is higher (0.962--0.966) at the cost of a slightly increased false discovery rate (FDR $\approx 0.068$--$0.070$), compared with the corresponding parametric stratum models (YBSP and YPSP), which show lower power (0.826--0.840) and modestly smaller FDR (0.046--0.053). 
Taken together, these results suggest that using mixed-effects BART in the outcome model is critical for accurate estimation and valid uncertainty quantification of heterogeneous CSACE, while a flexible stratum model further improves the ability to correctly identify always-survivors with acceptable control of false discoveries.

In summary, the fully nonparametric mixed-effects BART strategy (YBSB) provides the most favorable balance across all metrics, delivering the highest precision in CSACE estimation, near-nominal coverage, and strong principal stratum classification performance in this simulation design.

\begin{table}[htbp]
\centering
\caption{Simulation performance of four principal stratification estimators for CSACE and principal stratum classification.\label{tab:sim-summary}}
\begin{threeparttable}
\begin{tabular}{lcccccc}
\toprule
& PEHE & Abs.\ bias & Regret & Coverage  & Stratum power & Stratum FDR \\
\midrule
YBSB & 0.584 (0.088) & 0.448 (0.051) & 0.119 (0.020) & 0.951 (0.039) & 0.962 (0.021) & 0.068 (0.011) \\
YBSP & 0.586 (0.117) & 0.449 (0.070) & 0.120 (0.024) & 0.959 (0.044) & 0.826 (0.052) & 0.046 (0.014) \\
YPSB & 0.612 (0.062) & 0.505 (0.043) & 0.154 (0.019) & 0.447 (0.047) & 0.966 (0.019) & 0.070 (0.010) \\
YPSP & 0.649 (0.119) & 0.529 (0.083) & 0.156 (0.027) & 0.466 (0.051) & 0.840 (0.063) & 0.053 (0.012) \\
\bottomrule
\end{tabular}
\begin{tablenotes}
\footnotesize
\item Note: Entries are Monte Carlo means with standard deviations in parentheses, based on 100 simulated datasets. 
PEHE = precision in estimation of heterogeneous effect; Abs.\ bias = average absolute bias of estimated CSACE; 
Regret = expected regret for individualized treatment decisions; Coverage = empirical coverage probability of 95\% credible intervals; 
Stratum power and Stratum FDR summarize classification performance for the always-survivor principal stratum.
\end{tablenotes}
\end{threeparttable}
\end{table}

\subsection{Sensitivity to the nested MAR assumption}
\label{sec:sim-sensitivity}

We conducted a sensitivity analysis to examine the robustness of our methods when the nested MAR assumption in Section~\ref{sec:3.3} is violated. In particular, we considered a scenario in which missingness of survival status and the final outcome depends in part on individual-level covariates that are \emph{not} included in the analysis models.

We retained the same cluster structure, treatment assignment, principal-strata model, and outcome-generating mechanism as in the main simulation design. The only modification was to the missingness mechanism. Specifically, for each individual $(i,j)$ we generated two continuous, individual-level latent covariates
\(
U_{ij1}, U_{ij2} \,\overset{\text{i.i.d.}}{\sim}\, \mathcal N(0,1),
\)
which are not observed and are omitted from both the principal-strata and outcome models at the analysis stage. Survival-status recording and outcome-response indicators were then generated from logistic models that depend on both observed covariates and these unmeasured variables:
\[
\Pr\big(R_{ij}^S = 1 \mid Z_i, \mathbf X_{ij}, U_{ij1}, U_{ij2}\big)
= \operatorname{expit}\!\big(1.8 + 0.4 Z_i - 0.3 U_{ij1} + 0.5 U_{ij2} - 0.2 X_{ij3}\big),
\]
\[
\Pr\big(R_{ij}^Y = 1 \mid Z_i, \mathbf X_{ij}, U_{ij1}, U_{ij2}, S_{ij}^{\text{true}}=1, R_{ij}^S=1\big)
= \operatorname{expit}\!\big(2.5 + 0.5 Z_i - 0.2 U_{ij1} - 0.6 U_{ij2} + 0.3 X_{ij5}\big),
\]
where $S_{ij}^{\text{true}}$ denotes the true survival status under the assigned arm. By construction, $\{U_{ij1},U_{ij2}\}$ directly affect both $R_{ij}^S$ and $R_{ij}^Y$, but are excluded from the analysis models used by all four approaches (YBSB, YBSP, YPSB, and YPSP). Consequently, the nested MAR assumptions \eqref{ass:A1}-\eqref{ass:A2} are no longer satisfied, and the missingness mechanism is Missing Not At Random (MNAR) with respect to the analysis model.

For this sensitivity scenario, we applied exactly the same estimation procedures as in the main simulations, still assuming nested MAR and conditioning only on $(Z_i,\mathbf X_{ij})$ in the strata and outcome models. Performance was evaluated using the same metrics (PEHE, absolute bias, expected regret, empirical coverage, stratum power, and stratum FDR), allowing direct comparison with the main scenario and quantification of the impact of modest violations of nested MAR driven by unmeasured predictors of missingness.

\begin{table}[htbp]
\centering
\caption{Simulation performance of four principal stratification estimators when the nested MAR assumption does not hold.\label{tab:sim-sen}}
\begin{threeparttable}
\begin{tabular}{lcccccc}
\toprule
& PEHE & Abs.\ bias & Regret & Coverage  & Stratum power & Stratum FDR \\
\midrule
YBSB & 0.576 (0.090) & 0.441 (0.052) & 0.117 (0.020) & 0.954 (0.034) & 0.964 (0.017) & 0.072 (0.010) \\
YBSP & 0.604 (0.146) & 0.457 (0.085) & 0.120 (0.024) & 0.949 (0.060) & 0.825 (0.063) & 0.050 (0.014) \\
YPSB & 0.612 (0.077) & 0.503 (0.052) & 0.150 (0.016) & 0.440 (0.042) & 0.968 (0.016) & 0.074 (0.009) \\
YPSP & 0.640 (0.109) & 0.521 (0.072) & 0.154 (0.022) & 0.462 (0.044) & 0.849 (0.050) & 0.057 (0.012) \\
\bottomrule
\end{tabular}
\begin{tablenotes}
\footnotesize
\item Note: Entries are Monte Carlo means with standard deviations in parentheses, based on 100 simulated datasets generated under violation of the nested MAR assumption.
\end{tablenotes}
\end{threeparttable}
\end{table}

Table~\ref{tab:sim-sen} summarizes the sensitivity-analysis results. Overall, the relative ranking of the four approaches is unchanged: the proposed YBSB method continues to achieve the smallest PEHE and absolute bias and the lowest expected regret, together with the highest stratum power and low FDR, while the fully parametric YPSP approach performs worst across most metrics. Compared with the main simulation (Table~\ref{tab:sim-summary}), all methods experience some degradation in efficiency and coverage under MNAR, but the magnitude of deterioration for YBSB is modest. Taken together, these results suggest that our mixed-effects BART principal-stratification approach is reasonably robust to moderate violations of the nested MAR assumption of the type considered here.

\section{Demonstration of the WSD Telecare Questionnaire Study}

Posterior inference on principal strata composition indicated that, on average, 84.4\% of participants were always-survivors, 3.2\% were protected, and 12.4\% were never-survivors. To summarize the CSACE, we defined the set of \emph{likely always-survivors} as those participants with a posterior probability of at least 0.8 of belonging to the always-survivor stratum.

\begin{figure}[htbp]
  \centering
  \includegraphics[width=\textwidth]{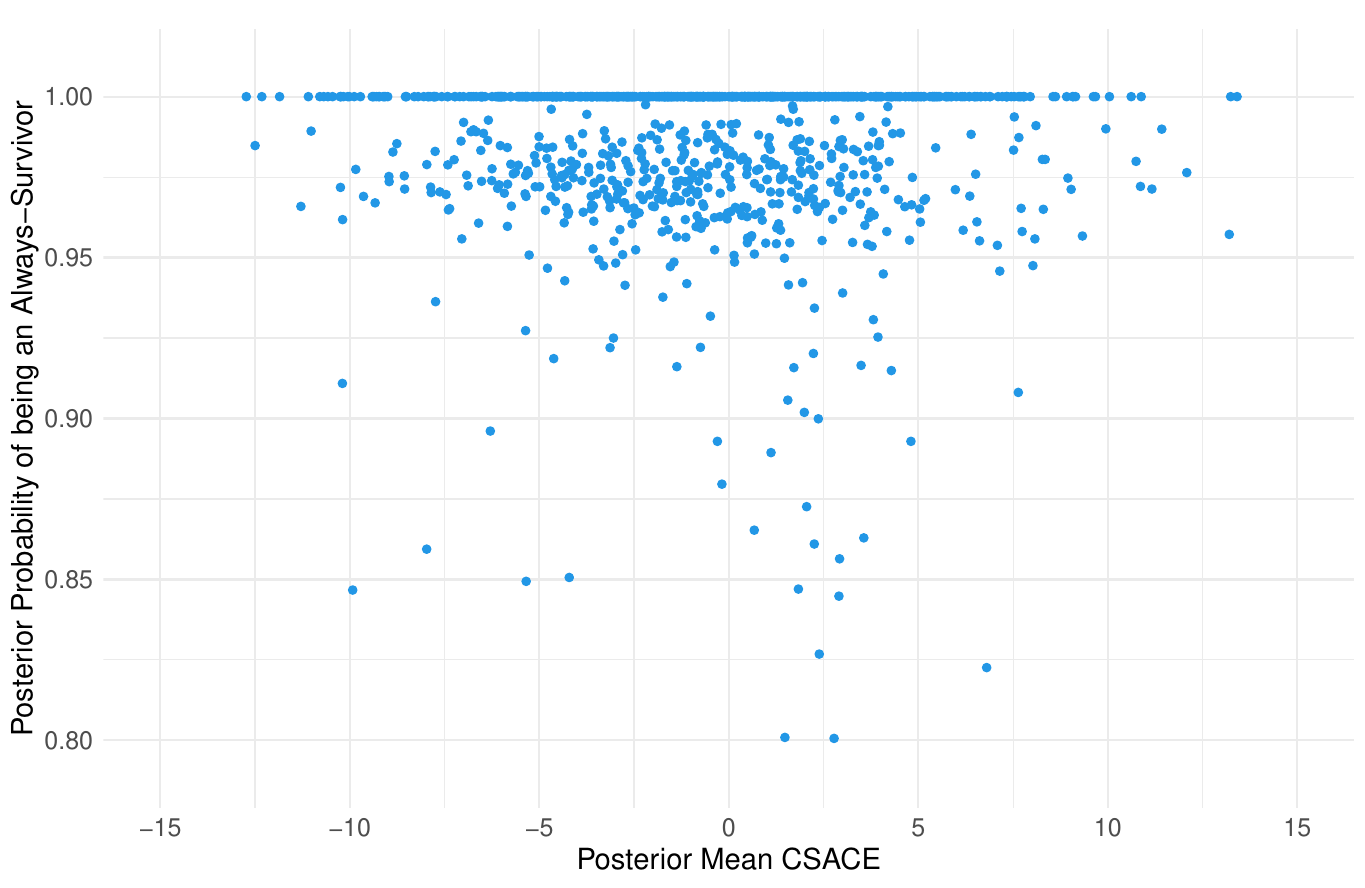}
  \caption{Posterior mean CSACE versus posterior probability of belonging to the always-survivor principal stratum for each participant. Higher values on the vertical axis indicate greater posterior certainty of always-survivor status; participants with probability at least 0.80 are treated as ``likely always-survivors'' in subsequent analyses.}
  \label{fig:scatter}
\end{figure}

Figure~\ref{fig:scatter} displays, for each participant, the posterior mean CSACE on the horizontal axis and the posterior probability of belonging to the always-survivor principal stratum on the vertical axis. We use a probability threshold of 0.80 to define ``likely always-survivors'', so points in the upper portion of the plot correspond to individuals with high posterior certainty of being in the always-survivor stratum. Most observations lie very close to 1.00 on the vertical axis, with the bulk of points above 0.95, suggesting high classification certainty for principal stratum membership. Only a small subset of participants have posterior probabilities below 0.90, and these are scattered across the range of CSACE values rather than concentrated at large positive or negative effects. Thus, classification uncertainty appears limited and does not show any obvious association with the magnitude or sign of the estimated CSACE.

Posterior summaries of the stratum proportions and the overall SACE are reported in Table~\ref{tab:posterior-summaries}.

\begin{table}[htbp]
\centering
\begin{threeparttable}
\caption{Results for the Survivor Average Causal Effect Estimate and the Proportion of Recipients in Each Principal Stratum}
\label{tab:posterior-summaries}
\begin{tabular}{lcc}
\toprule
\textbf{Quantity} & \textbf{Point estimate\tnote{a}} & \textbf{95\% CrI} \\
\midrule
Proportion of always-survivors & 0.844 & [0.799, 0.860] \\
Proportion of protected        & 0.032 & [0.000, 0.114] \\
Proportion of never-survivors  & 0.124 & [0.085, 0.140] \\
SACE                          & -0.568 & [$-3.580$, $2.532$] \\
\bottomrule
\end{tabular}
\begin{tablenotes}[flushleft]\footnotesize
\item[a] Point estimate is the posterior mean. CrI: equal-tailed 95\% credible interval.
\end{tablenotes}
\end{threeparttable}
\end{table}

\begin{figure}[htbp]
  \centering
  \includegraphics[width=\textwidth]{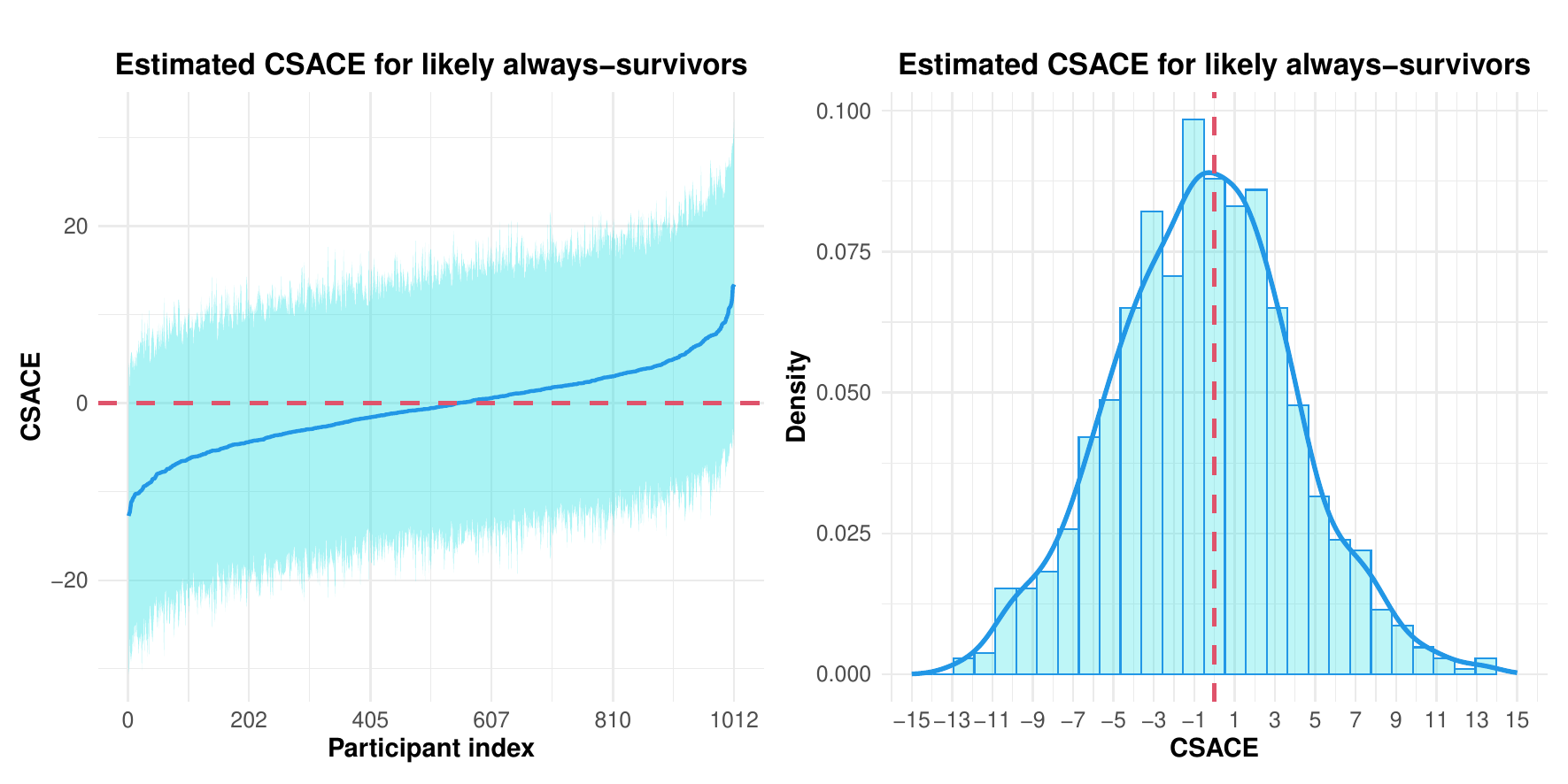}
  \caption{Distribution of CSACE on EQ-5D-VAS at 12 months among likely always-survivors. Left: participant-level posterior mean CSACEs (sorted) with 95\% credible intervals. Right: histogram of posterior mean CSACEs with a smoothed density curve.}
  \label{fig:csace_dist}
\end{figure}

Figure~\ref{fig:csace_dist} summarizes the distribution of individual CSACE estimates among likely always-survivors, defined as participants with posterior probability at least 0.80 of belonging to the always-survivor principal stratum. In the left panel, each position on the horizontal axis corresponds to a single participant, and the solid curve shows the posterior mean CSACE values sorted from smallest to largest. The light band around the curve represents 95\% credible intervals for each individual effect. Across patients, the posterior means range from modest negative to modest positive values, indicating meaningful heterogeneity in the estimated effects, but nearly all intervals include zero, suggesting limited evidence for large individual-level benefits or harms. The right panel displays a histogram of the posterior mean CSACEs with a smooth density overlay. The distribution is approximately symmetric and centered slightly below zero. The overall mean CSACE was $-0.57$ (95\% CrI $[-3.58,\,2.53]$), indicating no compelling difference between telecare and usual care in the 12 month EQ-5D-VAS summary score for the average always-survivor.

\begin{figure}[htbp]
  \centering
  \includegraphics[width=\textwidth]{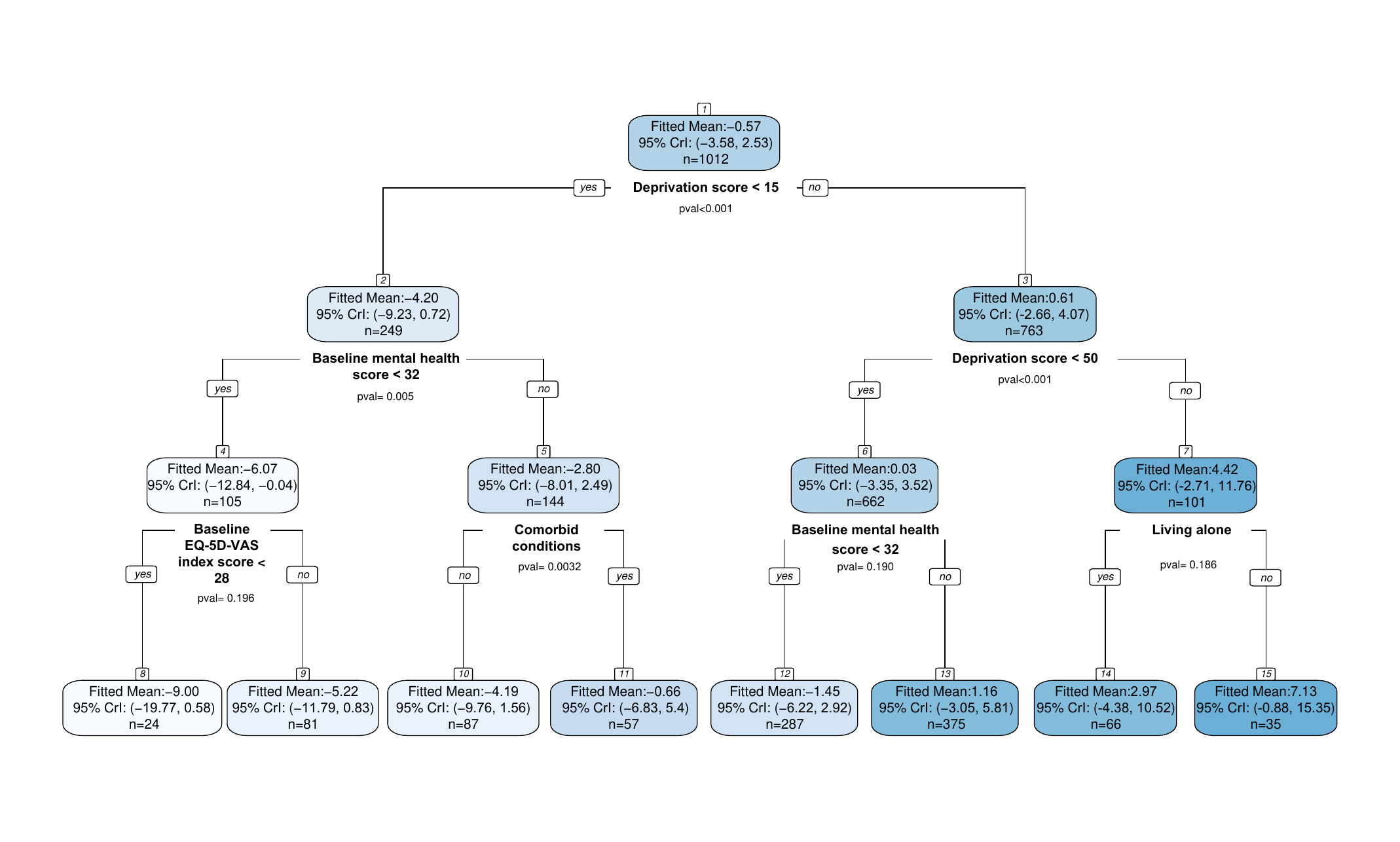}
  \caption{CART applied to posterior mean CSACEs for EQ-5D-VAS at 12 months among likely always-survivors. Nodes display posterior mean CSACE and 95\% CrI for each subgroup; splits are on baseline covariates.}
  \label{fig:cart_csace}
\end{figure}

To explore effect modification among likely always-survivors, we adopted a Bayesian “fit-the-fit” strategy: \cite{hahn2020bayesian} posterior mean CSACEs were used as responses in a classification and regression tree (CART) fitted to baseline covariates. The resulting tree (Figure~\ref{fig:cart_csace}) uses baseline deprivation score for the primary split at 15. Participants with deprivation $\ge 15$ had a mean CSACE of $0.61$ (95\% CrI $[-2.66,\,4.07]$), whereas those with deprivation $<15$ had a mean CSACE of $-4.20$ (95\% CrI $[-9.23,\,0.72]$), suggesting a greater tendency toward outcomes favoring usual care in the latter group.

Within the higher deprivation branch ($\ge 15$), a subsequent split at 50 yielded two distinct subgroups: deprivation $>50$ showed a larger mean CSACE of $4.42$ (95\% CrI $[-2.71,\,11.76]$), while deprivation $<50$ was near null at $0.03$ (95\% CrI $[-3.35,\,3.52]$). Among those most deprived (deprivation $>50$), living arrangement further refined the effect: participants not living alone exhibited the largest estimated benefit, with mean CSACE $7.13$ (95\% CrI $[-0.88,\,15.35]$). Within the moderate-deprivation range ($15 \le \text{deprivation} < 50$), baseline mental health emerged as the key moderator: mean CSACE $1.16$ (95\% CrI $[-3.05,\,5.81]$) for participants with good mental health (score $>32$) versus $-1.45$ (95\% CrI $[-6.22,\,2.92]$) for poor mental health (score $<32$).

Among participants with low deprivation ($<15$), the tree highlighted the joint role of baseline mental health and baseline EQ-5D-VAS. In particular, because higher scores indicate better baseline status, those with poor baseline mental health (score $<32$) and low baseline quality of life (EQ-5D-VAS $<28$) showed a mean CSACE of $-9.00$ (95\% CrI $[-19.77,\,0.58]$), favoring usual care in that subgroup. Overall, the tree points to higher deprivation—especially in non–alone households—as associated with greater benefit from telecare, whereas low deprivation combined with poorer mental health and lower baseline EQ-5D-VAS is associated with harm. Because many terminal node intervals include zero, these subgroup findings should be interpreted as exploratory.

\section{Discussion}

This study develops and demonstrates the application of machine learning tools in identifying treatment effect heterogeneity in a pragmatic gerontological trial. Specifically, we re-examine the WSD Telecare Questionnaire Study using a principal stratification causal framework for cluster-randomized trials with outcomes truncated by death. We embed mixed-effects BART to model strata membership and arm- and stratum-specific outcomes, yielding estimates of the SACE and individual survival treatment effect while accounting for cluster dependence and flexible nonlinear associations between baseline covariates and outcomes. Three findings are worth highlighting.

First, the mean CSACE on EQ-5D-VAS at 12 months was close to zero, indicating no compelling average improvement among always-survivors. Because SACE conditions on principal strata rather than observed survival, this estimand differs conceptually from intent-to-treat and survivor only analyses. Our result suggests that, for individuals who would survive regardless of arm, telecare did not yield apparent improvements in overall quality of life as captured by EQ-5D-VAS.

Second, the heterogeneity analysis among likely always-survivors points to clinically meaningful variation. The most important baseline features that appear to moderate the response to telecare include community deprivation, household structure, and mental health. Deprivation plausibly indexes environmental and social risks as well as unmet safety needs; when deprivation is higher, passive monitoring and escalation pathways have greater scope to prevent adverse events or provide reassurance, whereas in more advantaged settings the incremental value of monitoring may be limited and, for some, burdensome. Living arrangement modifies this gradient: in households with co-residents, reminders, help with equipment, interpretation of alerts, and faster response can translate monitoring into tangible benefit, while individuals living alone may not realize the same gains. Finally, poorer baseline mental health and low baseline EQ-5D-VAS mark greater symptom burden and psychosocial stress; passive monitoring alone may not address these drivers of low self-rated health and may heighten vigilance without relief, aligning with the negative signals observed in those branches. Taken together, these patterns argue for targeted deployment—prioritizing high deprivation recipients with household support while treating the subgroup signals as exploratory and requiring confirmation in pre-specified analyses. Meanwhile, it is worth highlighting that credible intervals at most terminal nodes include zero and subgroup sizes are modest; these findings should be regarded as hypothesis generating and framed as guidance for future, pre-specified confirmatory analyses rather than as definitive subgroup effects.

Third, the proposed estimation strategy handled multiple sources of incompleteness common in gerontology trials. The nested MAR assumption separates missing survival status from, conditional on observed survival, missing outcomes,
thereby permitting ignorable likelihood–based inference while retaining a clear estimand. Coupled with mixed–effects BART, the approach balances design based features of CRTs (cluster randomization, within–cluster dependence) with
nonparametric flexibility in the strata and outcome models, reducing vulnerability to functional form misspecification relative to linear mixed models.

Our results rely on a series of structural assumptions. At the trial design level, we adopt SUTVA with partial interference, meaning outcomes depend only on a person’s cluster assignment, with cluster-level randomization and positivity ensuring independence of assignment and potential outcomes; in the WSD setting, where randomization occurred at the practice level and telecare was delivered within households, these conditions are reasonable, although some unobserved cross-practice interactions cannot be ruled out. At the modeling layer, we use a principal stratification framework with monotonicity for survival ($S_{ij}(1)\ge S_{ij}(0)$), which reflects the intervention’s intent, accommodate outcomes truncated by death, and rely on principal exchangeability, an assumption that conditional on observed covariates, models for strata membership and outcomes are sufficiently rich to capture the relevant variation. While these conditions cannot be verified directly, they are standard in mixture-based analyses and provide a defensible path to identification. For missingness, we assume a nested MAR structure, where survival status and outcomes are missing at random given treatment and covariates, which is plausible given the WSD data and reasons for follow-up loss, though we note developing methods to relax this assumption would further strengthen robustness. Finally, we recognize that the analysis of effect modification is a post hoc exploration of potential mechanism rather than confirmatory. Nevertheless, the results highlight directions for future work and potential target population and improvement in the intervention design. Overall, our study shows that principal stratification combined with mixed effects BART yields inference that respects the trial design and is robust to nonlinear relations and higher order interactions. Substantively, the WSD reanalysis suggests that broad telecare deployment may not improve average quality of life among always-survivors; focusing implementation on participants in higher deprivation settings and those living with others may be more beneficial.

\section{Conclusion}

This study revisits a large, pragmatic telecare trial in older adults and asks a simple question that is often hard to answer: who is likely to benefit in terms of quality of life when some participants do not live to provide that
information? By focusing on people who would be alive at follow-up and by using a cluster-aware, flexible modeling approach suited to real world trial data, we were able to look beyond an overall average and examine how responses vary across patients.

Two messages stand out. First, on average, telecare did not improve the 12 month EQ-5D-VAS among people who would survive regardless of assignment. Second, there are meaningful differences across subgroups. Signals of benefit were more apparent for participants living in more deprived settings, especially those residing with others who could support consistent engagement with the system. In contrast, participants with lower deprivation combined with poorer baseline mental health and low baseline EQ-5D-VAS showed little evidence of benefit and may do better with usual care. These patterns are exploratory, but they suggest a practical strategy: target telecare to patients and households with greater capacity to translate monitoring into action.

From a translational perspective, several practical lessons emerge. Our analysis relies on a monotonicity assumption that telecare does not worsen survival compared with usual care. In many late-phase trials, where interventions have undergone safety piloting and are not expected to increase mortality, this is a reasonable working assumption; when there is genuine concern about treatment-induced harm, the framework can in principle be extended to include a ``harmed'' principal stratum, at the cost of substantially more complex mixture modeling and larger sample-size requirements to identify a typically rare subgroup. Likewise, the nested MAR assumption for missing survival status and quality-of-life outcomes is most credible when investigators collect rich baseline prognostic measures and understand the reasons for missingness in detail. In applications, we recommend that analysts (i) carefully document patterns and causes of missing data, (ii) assess the plausibility of nested MAR given the available covariates, and (iii) use simulation-based sensitivity analyses to gauge the impact of deviations from these assumptions. Finally, mixed-effects BART is particularly advantageous when there is a non-negligible intracluster correlation and a moderate to large number of clusters, whereas in settings with very small ICC or few clusters, simpler cluster-agnostic BART models may perform competitively and can be considered as a pragmatic alternative.

For clinicians and service planners, the implication is to move away from uniform deployment strategies. Prioritizing more deprived patients and ensuring household or caregiver support may yield greater gains, whereas those with substantial psychological burden and low baseline quality of life may need complementary services (e.g., mental health support, rehabilitation) alongside any monitoring technology. For researchers, future trials should prospectively evaluate such targeting rules, collect baseline factors that plausibly modify benefit, and pre-specify confirmatory subgroup analyses. The same analytical framework can be applied to other health-related quality of life outcomes collected in this study, including the SF-12 physical and mental component scores, enabling a broader view of patient-centered benefit. Replication in other settings, sensitivity analyses to key assumptions, and linkage to cost effectiveness will further clarify when and for whom telecare improves quality of life in later life.

\section*{Acknowledgments}

Research in this article was partially supported by the Patient-Centered Outcomes Research Institute\textsuperscript{\textregistered} (PCORI\textsuperscript{\textregistered} Award ME-2020C1-19220), the United States National Institutes of Health (NIH), National Heart, Lung, and Blood Institute (grant numbers R01-HL168202, R01-HL178513), as well as by the National Institute on Aging (NIA) of the National Institutes of Health (NIH) under Award Number U54AG063546-06A1, which funds NIA Imbedded Pragmatic Alzheimer’s Disease (AD) and AD-Related Dementias Clinical Trials Collaboratory (NIA IMPACT Collaboratory), the Yale OAIC (P30AG021342), and the Yale ADRC (P30AG066508).  All statements in this report, including its findings and conclusions, are solely those of the authors and do not necessarily represent the views of the NIH or PCORI\textsuperscript{\textregistered} or its Board of Governors or Methodology Committee. The Whole Systems ‘Demonstrator’ (WSD) telecare questionnaire study was funded by the Policy Research Programme in the Department of Health, UK.  The views expressed are not necessarily those of the Department. The WSD study was approved by Liverpool Research Ethics Committee (ref: 08/H1005/4). We acknowledge the generous sharing of the de-identified WSD telecare trial data from Drs Shashivadan P Hirani and Stanton P Newman at City, University of London, UK.

\section*{Conflict of Interest}
The authors have declared no conflict of interest.

\bibliographystyle{unsrtnat}
\bibliography{WSD}

\end{document}